# Mining Voter Behaviour and Confidence: A Rule-Based Analysis of the 2022 U.S. Elections

Md Al Jubair
*Computer Science and Engineering*
*Chittagong University of Engineering Technology*
Chittagong, Bangladesh
aljubairpollob@gmail.com

Mohammad Shamsul Arefin
*Computer Science and Engineering*
*Chittagong University of Engineering Technology*
Chittagong, Bangladesh
sarefin@cuet.ac.bd

Ahmed Wasif Reza
*Computer Science and Engineering*
*East West University*
Dhaka, Bangladesh
wasif@ewubd.edu

*Abstract* - **This study explores the relationship between voter trust and their experiences during elections by applying a rule-based data mining technique to the 2022 Survey of the Performance of American Elections (SPAE). Using the Apriori algorithm and setting parameters to capture meaningful associations (support ≥ 3%, confidence ≥ 60%, and lift > 1.5), the analysis revealed a strong connection between demographic attributes and various voting-related challenges, such as registration hurdles, accessibility issues, and queue times. For instance, respondents who indicated that accessing polling stations was "very easy" and who reported moderate confidence were found to be over six times more likely (lift = 6.12) to trust their county's election outcome and experience no registration issues. A further analysis, which adjusted the support threshold to 2%, specifically examined patterns among minority voters. It revealed that 98.16% of Black voters who reported easy access to polling locations also had smooth registration experiences. Additionally, those who had high confidence in the vote counting process were almost two times as likely to identify as Democratic Party. These findings point to the important role that enhancing voting access and offering targeted support can play in building trust in the electoral system, particularly among marginalized communities.**

## I. Introduction

Voting access disparities in the United States remain a critical issue, with mounting evidence suggesting that marginalized communities face systemic barriers to exercising their democratic rights. These disparities disproportionately affect racial minorities, low-income populations, and individuals with disabilities, raising concerns about the equity and inclusivity of the electoral process.

Traditional statistical methods, such as regression models, have been commonly employed to examine voting behaviour and access disparities. While effective for estimating correlations and trends, these models often fall short in revealing complex, multidimensional patterns that can offer deeper insights into how various demographic and situational factors intersect to create compounded obstacles.

Association Rule Mining (ARM) [2], a data-driven approach rooted in unsupervised learning, enables the discovery of hidden, interpretable patterns within large datasets. By analyzing frequent co-occurrences of attributes, ARM can illuminate nuanced relationships between voter characteristics and experienced barriers, facilitating more granular understanding and targeted policymaking.

In this study, we utilize Association Rule Mining (ARM) to analyze the 2022 Survey of the Performance of American Elections (SPAE) dataset [4], with the goal of identifying patterns that link voter trust to their demographic characteristics and election experiences such as registration issues, polling place accessibility, and wait times. Our contributions include the formulation of a focused analytical framework for electoral data, and the extraction of statistically significant and interpretable association rules.

The paper includes the following sections; section II contains literature review of related work. The proposed methodology is provided in section III. In Section IV a complete discussion is held on experimental result and we finalized the paper with a conclusion in Section V.

## II. Literature review

Over the past decade, there has been growing interest in understanding whether social media data can offer meaningful insights into electoral outcomes. While some studies suggest these platforms reflect public sentiment, others question their predictive reliability. One such comparative study [1] examined elections across India, Pakistan, and Malaysia, using techniques like tweet volume analysis, sentiment-based modeling, and social network analysis. Although the research covered multiple countries, its findings were more conclusive for India and Pakistan. The authors highlight key limitations, noting that despite the use of advanced models, relying solely on social media may not provide accurate forecasts. The broader question whether digital sentiment can truly capture voter behavior remains open.

One study in Portugal by Carvalho and colleagues [9] found that young people reacted more strongly to negative news than to positive news. Bad news made them think worse of the government, and it also pushed undecided voters to stay home or support the opposition. In another study comparing seven democratic countries [10], researchers saw that voters in places like the UK, Australia, and Canada were more likely to re-elect their leaders than voters in the U.S. or France. The influence of a leader's reputation or the opinions of family and friends may account for this difference. In the UK, Horan et al. [20] looked at voting in the 2019 election and found that local factors like education, health, and ethnic background played a big role in how people voted. They also discovered that people living near each other often voted in similar ways. A study in Australia [11] looked at how young voters decide who to vote for. It showed that people who had voted before were more likely to look for information and stay committed to voting, even if they didn't always know a lot about politics.

In parallel, another body of research has shed light on long-standing disparities in voting access and participation, especially along racial and socioeconomic lines. Work from organizations like the Brennan Center for Justice and the MIT Election Lab [4] has revealed that Black and Hispanic voters often encounter more obstacles at the polls ranging from longer wait times to registration issues and inadequate polling resources. Other demographic factors such as income, education, geographic location, and disability status have also been shown to affect electoral participation. Rural voters, for instance, may have to travel farther to reach polling places, while urban residents in minority-dominated neighborhoods frequently face overcrowded or under-resourced voting facilities.

Previous studies have used simple statistics to show how things like race, income, or location affect a person's ability to vote. While that can help show broad trends, it often misses the deeper story, like how several issues combine to affect someone's experience. Given these layered influences and the limits of traditional analysis, more adaptable techniques are needed to uncover the deeper patterns behind voter trust. To get a clearer understanding, we focused on real voter survey data from the 2022 election and used a method called Association Rule Mining. It assists in identifying patterns that may not always be readily apparent, such as trust in the electoral process. While common methods can show overall trends, they sometimes miss how these different pieces fit together. Our approach lets us look beyond the surface and pick up on details that standard tools might overlook.

### III. METHODOLOGY

There are several steps for the study are presented in this section. To explore patterns in voter behavior, data from the 2022 Survey of the Performance of American Elections (SPAE) was utilized. The initial step involved bringing in the raw survey responses and refining them by addressing incomplete entries and standardizing the format. After this, attention turned to identifying the most relevant details factors like how people voted, their background characteristics, and their experiences during the voting process. With this curated set of information, pattern-finding techniques such as Apriori [2] were applied to uncover recurring combinations of responses. These results were then carefully reviewed to filter out noise and focus on the most meaningful connections. What emerged was a set of practical insights that offered a clearer picture of voting trends and behaviours, useful for shaping future improvements in the electoral system. To achieve the goal, we have sketched a framework Fig.1.

#### A. Dataset Overview

We use the 2022 Survey of the Performance of American Elections (SPAE) [4], a comprehensive national survey of registered voters conducted post-election. The dataset contains over 10,200 responses and includes variables related to voting experiences, access barriers, demographic details, and voting methods.

#### B. Limitation of the Dataset

The SPAE 2022 [4] offers useful insights from across the country, but there are few limitations:

a) **Sampling Bias:** The survey exclusively questioned individuals who had already registered to vote, resulting in a total of approximately 10,200 participants. That means anyone who wasn't registered or who gave up trying to register wasn't included. Because of this, some voting issues might appear less common than they actually are. Additionally, the survey relies on individuals' self-reports of their voting behaviour's which haven't undergone a thorough verification against official voting records.

b) **Demographic Imbalance:** The goal was to collect about 200 answers from each state; some groups may still be under-represented. For instance, the survey may not fully capture Native American voters residing in rural areas. Additionally, there's a political skew among respondents: nearly half identified as Democrats (47.9%), while just 37% said they were Republicans.

c) **Response Bias:** Many of the questions asked people to recall experiences or feelings, like how long they waited in line or how sure they were about the vote counting. Memories and feelings can be unreliable. Since the survey was conducted online and by phone, people without steady internet or phone access, such as some older adults or low-income voters, might have been left out.

#### C. Data Preparation

Missing or inconsistent responses were treated with careful imputation or exclusion. For instance, question `q2` was set to NaN for voters who skipped it, and logical consistency checks were performed to remove contradictory

responses (e.g., someone reporting both mail and in-person voting)

### D. Feature Selection

From the initial set of 187 variables, we retained 15 that were directly relevant to the analysis based on the prior studies [3]. These included:

a) **Demographics:** race, income, age, gender, disability status (q55), voter location
b) **Voting experience:** wait times (q12), polling place accessibility issues (q5), registration difficulties (q9), vote counting confidence on your vote (q39), vote counting confidence on the county or city(q40), vote counting confidence on the state (q41), problem with voter registration (q9), line length to vote (q12),
c) **Political interest:** (newsint)
d) **Voting method:** mail, early, and in-person voting (q4)

We binned continuous variables such as age into ranges (18–29, 30–44, 45–64, 65+) to facilitate rule mining.

### E. Association Rule Mining Algorithm

We employed the Apriori algorithm [12] to generate frequent item sets and derives association rules. Additionally, we selected it because of its interpretability, transparency, and suitability for policy-oriented analysis, where explaining the relationships between variables is more important than optimizing prediction performance. Pseudocode 1 explains the Apriori algorithm.

---

**PSEUDOCODE 1: APRIORI ALGORITHM**

*Input:* transactions = list of all transactions, min_support = minimum support threshold
*Output:* all_frequent_itemsets — list of all frequent itemsets

1  **frequent_itemsets** ← find all frequent 1-itemsets in transactions
2  **all_frequent_itemsets** ← frequent_itemsets
3  **itemset_size** ← 2
4  **while frequent_itemsets is not empty:**
5      **candidate_itemsets** ← generate new itemsets of size itemset_size from frequent_itemsets
6      **count_map** ← empty map to count support
7      **for each transaction in transactions:**
8          **for each candidate in candidate_itemsets:**
9              **if candidate is a subset of transaction:**
10                 increment count_map[candidate] by 1
11     **frequent_itemsets** ← candidates from count_map with support ≥ min_support
12     add frequent_itemsets to all_frequent_itemsets
13     **itemset_size** ← itemset_size + 1
14 **return all_frequent_itemsets**

---

### F. Rule Filtering and Interpretation

To focus on impactful rules, we applied the following thresholds based on domain-specific precedents.

TABLE I
ASSOCIATION RULE MINING THRESHOLD PARAMETERS

| Parameter | Value | Description |
|---|---|---|
| Minimum support | 0.03 | Rule must appear in at least 3% of transactions [3] |
| Minimum confidence | 0.60 | Rule must hold true in ≥60% of applicable cases [2] |
| Minimum lift | 1.50 | Rule must be ≥1.5× more likely than random chance [5] |

## IV. EXPERIMENTAL RESULTS AND DISCUSSION

We discovered 14,106 equity-related association rules. The top rules revealed striking patterns in the table II.

These results underscore strong co-occurrence between positive voting confidence, ease of voting, and being White, suggesting disparities for other racial groups. High lift values (often exceeding 6.0) indicate meaningful deviations from random chance and affirm the presence of systemic voting inequities. We started by applying a support threshold of 3%, which allowed for the identification of several rules with strong confidence and interpretive relevance. These rules bring light on how different aspects of the voting experience are interrelated. A second phase of the analysis, using a slightly reduced support threshold of 2%, was conducted with a specific lens on minority voter patterns most notably among Black respondents to ensure nuanced trends weren't overlooked.

### A. Observations on Voting Equity

One of the clearest and most repeated patterns from the data is the close connection between voter confidence at the local and state levels. Voters who reported low trust in their county's vote counting process often carried that doubt to expressing a similar lack of faith in statewide outcomes. This relationship is especially visible in a rule with a lift value of 8.31, suggesting a strong dependency between trust in different layers of the electoral system. Such findings imply that voter confidence may cascade issues at the local level can potentially damage the perceived credibility of the system overall. Ease of access to polling stations also emerged as a recurring theme. Across multiple rule combinations, voters who described their voting experience as "very easy" and had at least moderate confidence in vote counting were also more likely to report no issues with registration and to trust county-level election results. A notable example includes a rule with a lift of 6.12, indicating that accessible polling locations when paired with even a modest degree of confidence are linked with higher satisfaction and trust. These patterns point toward intersecting structural barriers affecting already marginalized populations. The intersectionality captured by multi-attribute rules reveals how factors such as race, income, and disability compound one another. Policymakers and electoral authorities should prioritize accessibility enhancements and equitable resource allocation in rural and underserved urban areas. Specific actions could include deploying mobile polling units, increasing early voting availability, and reforming voter registration procedures. Top equity rules are visualized in the figure 2.

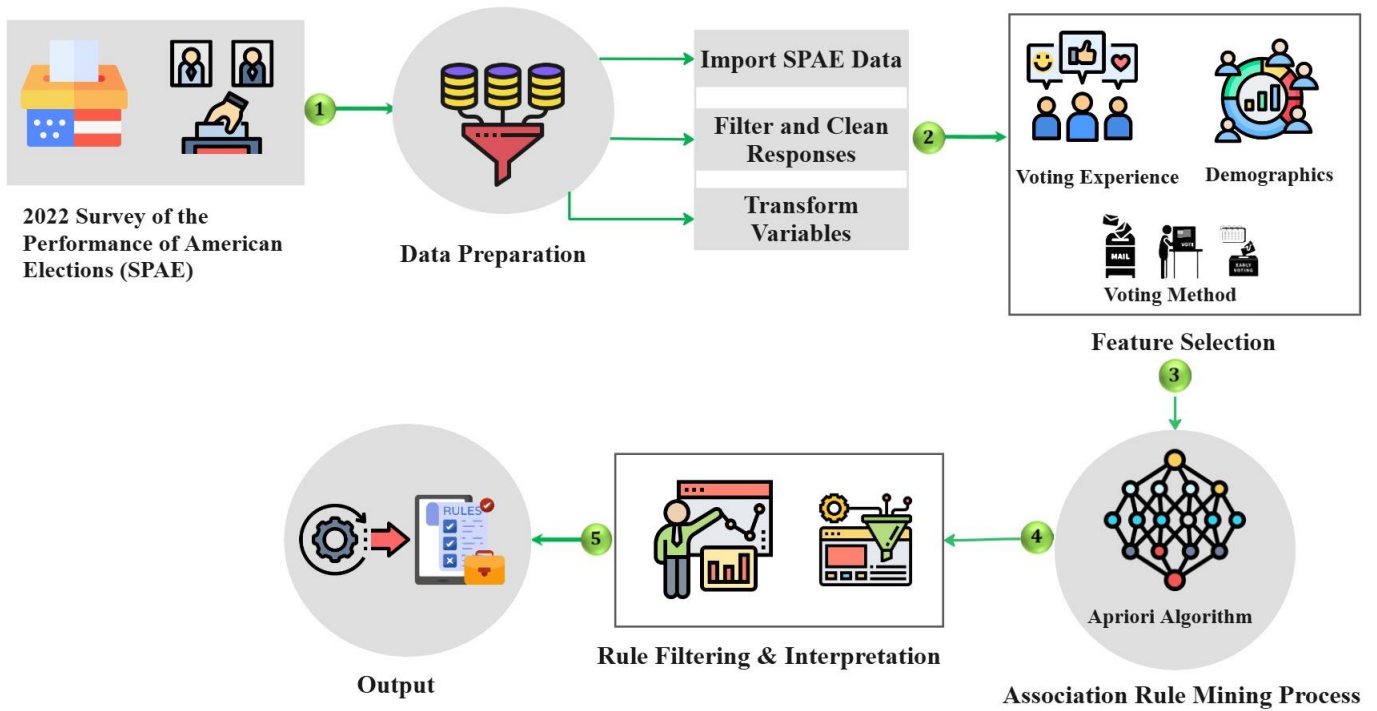

Fig. 1. The framework of the proposed study.

TABLE II
TOP ASSOCIATION RULES FROM ELECTION DATA ANALYSIS

| | Serial | Antecedent (IF) | Consequent (THEN) | Interpretation | Support (%) | Confidence (%) | Lift |
|---|---|---|---|---|---|---|---|
| | | | | **Minimum Support = 0.03** | | | |
| **Top 10 Voting Equity Rules** | E1 | 'q40_Not too confident' | 'q41_Not too confident' | Voters who doubt local vote counting are 8.3 times more likely to distrust state results. | 3.44 | 61.4 | 8.31 |
| | E2 | 'q12_Not at all', 'q39_Somewhat confident' | 'q5_Very easy', 'q40_Somewhat confident' | Voters with moderate confidence and no wait times are likely to report easy access and county-level trust. | 3.17 | .70.4 | 6.83 |
| | E3 | 'q5_Very easy', 'q39_Somewhat confident' | 'q40_Somewhat confident', 'q9_No' | Easy access and moderate confidence make voters 6 times more likely to trust county results and avoid registration issues. | 6.36 | 76.8 | 6.12 |
| | E4 | q4_Voted in person on Election Day (at a polling place or precinct)', 'q39_Somewhat confident' | 'q40_Somewhat confident', 'q9_No' | In-person voters with moderate confidence rarely face registration issues but still doubt county results. | 5.86 | 76.8 | 6.12 |
| | E5 | 'q40_Somewhat confident', 'q12_Not at all' | 'q9_No', 'q39_Somewhat confident' | Voters who trust county results and face no wait times are 6.1 times more likely to have smooth registration and personal vote counting confidence. | 3.45 | 62.4 | 6.09 |
| | E6 | 'q12_Not at all', 'q39_Somewhat confident' | 'q40_Somewhat confident', 'q9_No' | Voters with no lines and moderate confidence are 6.1 times more likely to trust county results and avoid registration issues. | 3.45 | 76.5 | 6.09 |
| | E7 | 'q9_No', 'q39_Somewhat confident' | 'q5_Very easy', 'q40_Somewhat confident' | Smooth registration and moderate confidence make voters 6 times more likely to report easy access and trust county elections. | 6.36 | 62.16 | 6.03 |
| | E8 | 'q5_Very easy', 'q40_Somewhat confident' | 'q9_No', 'q39_Somewhat confident' | Easy access and county trust lead to 6 times higher hassle-free registration and voting confidence. | 6.36 | 61.75 | 6.03 |
| | E9 | 'q4_Voted in person on Election Day (at a polling place or precinct)', 'q40_Somewhat confident' | 'q9_No', 'q39_Somewhat confident' | In-person voters trusting county results are 6 times more likely to experience smooth registration and personal vote confidence. | 5.86 | 61.39 | 5.99 |
| | E10 | 'q40_Somewhat confident', 'q12_Not at all' | 'q41_Somewhat confident', 'q5_Very easy' | County confidence and no wait times make voters 5.3 times more likely to find polling places accessible and trust statewide results. | 3.66 | 66.31 | 5.29 |
| | | | Minimum Support = 0.02 | | | | |
| **Top 5 Minority Specific Rules** | M1 | q39_Very confident, race_Black | pid3_Democrat | Black voters confident in their vote count are nearly twice as likely to identify as Democrats. | 3.2 | 74.65 | 1.99 |
| | M2 | race_Black, q41_Very confident | pid3_Democrat | Black voters very confident in state-level vote counting have a 74.2% chance of identifying as Democrats. | 3.12 | 74.18 | 1.98 |
| | M3 | q40_Very confident, race_Black | pid3_Democrat | Black voters very confident in county-level vote counting have a 73.9% chance of identifying as Democrats. | 3.36 | 73.92 | 1.97 |
| | M4 | gender_Female, race_Black | pid3_Democrat | Black women identify as Democrats at 70.7% rate. | 3.58 | 70.65 | 1.88 |
| | M5 | q5_Very easy, race_Black | q9_No | When polling places are accessible, 98.2% of Black voters report no registration issues. | 3.14 | 98.16 | 1.87 |

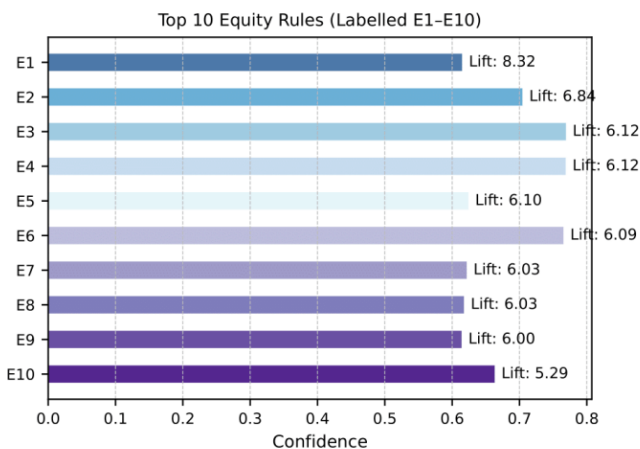

Fig. 2. Visualization of top 10 rules for Equity.

### B. Patterns Among Minority Group

When attention was directed toward patterns involving minority voters particularly Black participants a few consistent trends emerged. Across multiple rules, those who were "very confident" that their personal, county, or statewide votes were accurately counted were found to be roughly twice as likely to identify with the Democratic Party. In one such rule, the confidence stood at 74.65%, with a lift value close to 2. These findings align with known political leanings but add quantitative depth to our understanding of the relationship between trust in the electoral system and political identity among Black voters.

Another notable finding concerns voter registration. When Black voters reported that their polling place was "very easy" to access, nearly all over 98% said they encountered no registration issues. This rule is not only significant in terms of public policy, but it is also a high confidence rule. It implies that administrative complexity can be significantly decreased by improving polling station access logistics, especially for historically under-represented populations. In the figure 3, top minority rules are visible.

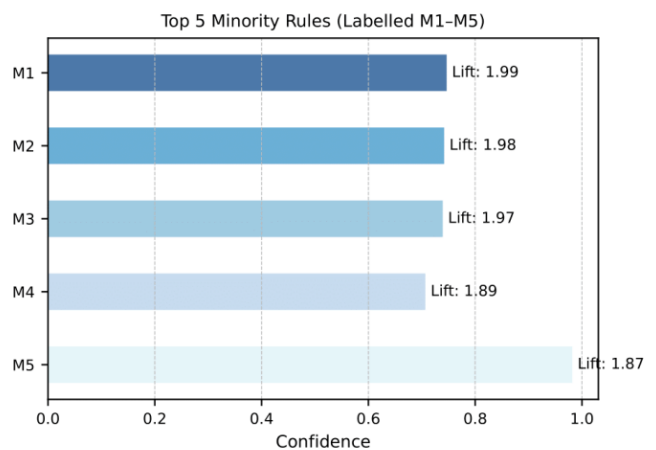

Fig. 3. Visualization of top 5 rules for Minority.

## C. Comparative Analysis

We applied both Apriori [12] and FP-Growth [6] on our dataset to see what kinds of rules they'd pull out. Apriori gave us a large number over 17,000 rules while FP-Growth returned only 82. Since we were mainly interested in fairness-related outcomes, we dug deeper into equity and minority-related rules. Apriori turned up around 14,000 equity-focused rules and about 40 connected to minority cases. In contrast, FP-Growth only found 68 equity rules and none for minorities. Interestingly, even with such a big difference in rule count, the confidence scores from both were pretty close. Apriori just seemed better at spotting those less frequent, but often more meaningful, patterns. The side-by-side breakdown is shown in Table III.

TABLE III
PERFORMANCE COMPARISON OF ASSOCIATION RULE MINING ALGORITHMS

| Algorithm | Rules | Equity Rules | Minority Rules | Avg Support | Avg Confidence | Avg Lift | Time (s) |
|---|---|---|---|---|---|---|---|
| Apriori [12] | 17,337 | 14,106 | 40 | 0.089 | 0.76 | 1.395 | 1.6078 |
| FP-Growth [6] | 82 | 68 | 0 | 0.45 | 0.79 | 1.303 | 1.5256 |

## V. CONCLUSION AND FUTURE WORK

In this work, we explored the 2022 SPAE dataset using association rule mining to investigate patterns in voting access and confidence across different demographic groups. A few findings stood out. For example, 98.16% of Black voters who reported easy polling access experienced no registration issues, and those with high confidence in vote counting were nearly twice as likely to identify as Democrats.

The method worked well for surfacing patterns, but it's worth noting that outcomes depend heavily on where support and confidence thresholds are set. Future efforts should take a more systematic route, ideally through sensitivity analysis, to determine the most meaningful settings.

We also see value in trying other techniques like Eclat or even combining ARM with machine learning to catch more subtle patterns. There's room to go further, too, by layering in time-based trends, spatial variation, or even qualitative perspectives. Together, these could help uncover deeper insights and inform more responsive voting policies, particularly for groups that often face systemic barriers.